\documentclass{llncs}
\usepackage{amsfonts}
\usepackage{latexsym,amsmath,epsfig,amssymb,graphics,amscd,amsfonts}

\newcommand{\oomit}[1]{}

\newcommand{\QQ}{\mathbb{Q}}
\newcommand{\ZZ}{\mathbb{Z}}
\newcommand{\RR}{\mathbb{R}}

\newcommand{\ii}{{\rm\bf i}}

\newcommand{\sm}[1]{{\small ${#1}$}}

\newcommand{\Pone}{${\tt P_1}$}

\newtheorem{notation}{Notation}
\date{}
\title{ Termination of Linear Programs with Nonlinear Constraints
\thanks{This work is supported in part by NKBRPC-2004CB318003,
NSFC-60573007, NSFC-90718041 and NKBRPC-2005CB321902.} }
\author{Bican Xia, Zhihai Zhang}
\institute{ School of Mathematical Sciences, Peking University\\
xbc@math.pku.edu.cn, infzzh@gmail.com }

\begin{document}
\label{firstpage}
\maketitle

\begin{abstract}
In \cite{Tiw04} Tiwari proved that termination of linear programs
(loops with linear loop conditions and updates) over the reals is
decidable through Jordan forms and eigenvectors computation. In
\cite{Mark06} Braverman proved that it is also decidable over the
integers. In this paper, we consider the termination of loops with
polynomial loop conditions and linear updates over the reals and
integers. First, we prove that the termination of such loops over
the integers is undecidable. Second, with an assumption, we provide
an algorithm to decide the termination of a class of such programs
over the reals. Our method is similar to that of Tiwari in spirit
but uses different techniques. Finally, we conjecture that the
termination of linear programs with polynomial loop conditions over
the reals is undecidable in general by 
reducing the problem to another decision problem related to number
theory and ergodic theory, which we guess undecidable.
\end{abstract}

\section{Introduction}
Termination analysis is an important aspect of program verification.
Guaranteed termination of program loops is necessary for many
applications, especially those for which unexpected behavior can be
catastrophic. For a generic loop
\[{\bf while}\quad
(conditions)\quad \{updates\}, \] it is well known that the
termination problem is undecidable in general, even for a simple
class of polynomial programs \cite{BMS}. In \cite{Blon} Blondel et
al. proved that, even when all the conditions and updates are given
as piecewise linear functions, the termination of the loop remains
undecidable.

In \cite{Tiw04} Tiwari proved that termination of the following
programs is decidable over $\mathbb{R}$ (the real numbers)
\[{\tt P_0:}\quad\quad {\bf while}\quad
(BX>b)\quad \{X:=AX\}, \] where $A$ and $B$ are respectively
$n\times n$ and $m\times n$ matrices, $BX>b$ represents a
conjunction of linear inequalities over the state variables $X$ and
$X:=AX$ represents a (deterministic) linear update of each variable.
Subsequently in \cite{Mark06} Braverman proved that the termination
of ${\tt P_0}$ is decidable over $\mathbb{Z}$ (the integers).

In this paper, we consider the problem of termination of the
following loop:
\[{\tt P_1:}\quad\quad{\bf while}\quad
(P(X)>0)\quad \{X:=AX\}, \] where $X=[x_1\ \ldots\ x_N]^{\rm T}$ is
the vector of state variables of the program, $P(X)=[P_1(X)\
P_2(X)\, \ldots\, P_m(X)]^{\rm T}$ are polynomial constraints, each
$P_i(X)$ $(1\le i\le m)$ is a polynomial in $\mathbb{Q}[X]$ and $A$
is an $N\times N$ matrix over $\mathbb{Q}$ (the rational numbers).
That is to say, we replace the linear constraints in ${\tt P_0}$
with polynomial constraints and keep linear updates unchanged.

There are some well known techniques for deciding termination of
some special kinds of programs. Ranking functions are most often
used for this purpose. A ranking function for a loop maps the values
of the loop variables to a well-founded domain; further, the values
of the map decrease on each iteration. A linear ranking function is
a ranking function that is a linear combination of the loop
variables and constants. Recently, the synthesis of ranking
functions draws increasing attention, and some heuristics concerning
how to automatically generate linear ranking functions for linear
programs have been proposed in \cite{cs01,DGG00,PR}. In \cite{PR}
Podelski et al. provided an efficient and complete synthesis method
based on linear programming to
construct linear ranking functions. 
In \cite{Chen} Chen et al. proposed a method to generate non-linear
ranking functions based on semi-algebraic system solving. However,
existence of ranking function is only a sufficient condition on the
termination of a program. It is not difficult to construct programs
that terminate, but do not have ranking functions.

To solve the problem of termination of ${\tt P_1}$, we do not use
the technique of ranking functions. Our method is similar to that of
Tiwari in spirit. Our main contributions in this paper are as
follows. First, we prove that the termination of ${\tt P_1}$ over
$\mathbb{Z}$ is undecidable. Then it is easy to prove that, if
``$>$" is replaced with ``$\ge$" in ${\tt P_1}$, termination of the
resulted ${\tt P_1}$ over $\mathbb{Z}$ is undecidable either.
Second, with an assumption, we provide an algorithm to decide the
termination of ${\tt P_1}$ over $\mathbb{R}$. Finally, we conjecture
that the termination of ${\tt P_1}$ over $\mathbb{R}$ is undecidable
in general by constructing a loop and reducing the problem to
another decision problem related to number theory and ergodic
theory, which we guess undecidable.

The rest of the paper is organized as follows. Section 2 proves the
undecidability of ${\tt P_1}$ and its variation over $\mathbb{Z}$.
Section 3 introduces our main algorithm. The main steps of the
algorithm are outlined first and some details of the steps are
introduced separately in several subsections. With an assumption, we
prove the correctness of our algorithm at the end of Section 3. 
After presenting our conjecture that the termination of ${\tt P_1}$
is generally undecidable in Section 4, we conclude the paper in
Section 5.

\section{Undecidability of ${\tt P_1}$ over $\mathbb{Z}$ }
\begin{definition}
A loop with $N$ variables is called {\em terminating} over a ring
$R$ if for all the inputs $X\in R^N$, it is terminating; otherwise
it is called {\em nonterminating}.
\end{definition}

The undecidability of ${\tt P_1}$ is obtained by a reduction to
Hilbert's $10^{th}$ problem. Consider the following loop:{\small
\[{\tt P_2:} \quad\quad{\bf while} \quad
(x_{m+1}-f(x_1,\ldots,x_m)^2>0)\quad \{X:=AX\}\]} where
$X=\begin{bmatrix}x_1&\ldots&x_{m+1}\end{bmatrix}^{\rm T}$, $A={\rm
diag}(1,...,1,1/2)$ is a diagonal matrix and $f(x_1,\ldots,x_m)$ is
a polynomial with integer coefficients.

\begin{lemma}\label{LE:reduction}
For any input $(x_1,\ldots,x_{m+1})\in \mathbb{Z}^{m+1}$, ${\tt
P_2}$ terminates if and only if $f(x_1,\ldots,x_m)$ does not have
integer roots.
\end{lemma}
\begin{proof} $(\Rightarrow)$ If \sm{f(x_1,\ldots,x_m)} has an integer
root, say \sm{(y_1,\ldots,y_m)}, obviously \sm{{\tt P_1}} does not
terminate with the input \sm{Y=(y_1,\ldots,y_m,1)}.

$(\Leftarrow)$ If \sm{f(x_1,\ldots,x_m)} has no integer roots, for
any given \sm{X\in \mathbb{Z}^{m+1}}, \sm{-f(x_1,\ldots,x_m)^2} is a
fixed negative number. Because \sm{(x_1,...,x_m)} will never be
changed and \sm{(1/2)^n\rightarrow 0} as \sm{n\rightarrow +\infty},
the loop will terminates after sufficiently large \sm{n} iterations.
\end{proof}

\begin{theorem}\label{g}
Termination of ${\tt P_1}$ over $\mathbb{Z}$ is undecidable.
\end{theorem}
\begin{proof} Because the existence of an integer root of an arbitrary
Diophantine equation is undecidable, the termination of ${\tt P_1}$
over $\mathbb{Z}$ is undecidable according to Lemma
\ref{LE:reduction}.
\end{proof}

If ``$>$" is substituted with ``$\ge$" in ${\tt P_1}$, the loop
becomes{\small
\[{\tt P_1^\prime:}\quad\quad {\bf while}\quad
(P(X)\ge 0)\quad \{X:=AX\}. \]} Analogously, we denote by ${\tt
P_2^\prime}$ the loop obtained by substituting ``$\ge$" for ``$>$"
in ${\tt P_2}$. It is easy to see that Lemma \ref{LE:reduction}
still holds for ${\tt P_2^\prime}$. Then we get the following
theorem.

\begin{theorem}\label{ge}
Termination of ${\tt P_1^\prime}$ over $\ZZ$ is undecidable.
\end{theorem}

\section{Relatively Complete Algorithm for Termination of ${\tt P_1}$ over $\mathbb{R}$}

To decide whether \Pone\ terminates on a given input $X\in \RR^N$,
it is natural to consider a general expression of $A^nX$, for
instance, a unified formula expressing $A^nX$ for any $n$. If one
has such unified formula of $A^nX$, one can express the values of
$P(X)$ after $n$ iterations. Then, for each element of $P(X)$ (each
constraint), {\it i.e.}, $P_i(X)$, one may try to determine whether
$P_i(X)>0$ as $n\rightarrow +\infty$ by guessing the dominant term
of $P_i(X)$ and deciding the sign of the term.

That is the main idea of our algorithm which will be described
formally in Subsection \ref{sec:main}. At several main steps of our
algorithm, a few techniques and results in number theory and ergodic
theory are needed. For the sake of clarity, the details are
introduced subsequently in the next subsections.

The first subsection is devoted to expressing $A^nX$ in a unified
formula.


\subsection{General Expression for $A^nX$ by Generating
Function}\label{sec:An}

In this subsection the general expression of $A^nX$ will be deduced
with generating function, not Jordan form.
\begin{lemma} {\rm \cite{Stan97}}\label{Generating function}
Let $\alpha_1,\ldots,\alpha_d$ be a sequence of complex numbers,
$d\ge1$ and $\alpha_d\ne 0$. The following conditions on a function
$f:\mathbb{N}\rightarrow\mathbb{C}$ are equivalent to each other:
\begin{description}
\item[i.] $\sum\limits_{n\ge0}f(n)x^n=\frac{P(x)}{Q(x)}$, where
$Q(x)=1+\alpha_1x+\ldots+\alpha_dx^d$ and $P(x)$ is a polynomial in
x of degree less than d.
\item[ii.] For all $n\ge0$,
$f(n+d)+\alpha_1f(n+d-1)+\alpha_2f(n+d-2)+\ldots+\alpha_df(n)=0$.
\item[iii] For all $n\ge 0$,
$f(n)=\sum\limits_{i=1}^{k}P_i(n)\gamma_i^n$, and
$Q(x)=1+\alpha_1x+\alpha_2x^2+\ldots+\alpha_dx^d=\prod\limits_{i=1}^{k}(1-\gamma_ix)^{d_i},$
where the $\gamma_i$'s are distinct, and $P_i(n)$ is a polynomial in
n of degree less than $d_i$.
 \end{description}
\end{lemma}


\begin{corollary} \label{co:fn}
$A$ is a $d\times d$ square matrix with its entries in $\mathbb{Q}$.
Suppose that the characteristic polynomial of $A$ is
$D(x)=x^d+\alpha_1x^{d-1}+\ldots+\alpha_{d-s}x^s,$ where
$\alpha_{d-s}\ne0$ and $s\ge 0$. Define $f(n)=A^{n+s}X$ and let
$f_j(n)$ be the $j$-th component of $f(n)$. Then for each $j$,
$f_j(n)$ can be expressed as{\small
\begin{equation}\label{eq:fn}
f_j(n)=\sum\limits_{i=1}^{k}p_{ji}(n)\xi_i^n,
\end{equation} }
where $\xi_i$'s are all the distinct nonzero complex eigenvalues of
$A$ and $p_{ji}(n)$ is a polynomial in $n$ of degree less than the
multiplicity of $\xi_i$.
\end{corollary}
\begin{proof}
First, \sm{A^d+\alpha_1A^{d-1}+\cdots+\alpha_{d-s}A^s=0} because
\sm{D(x)} is the characteristic polynomial of \sm{A}. So, for any
\sm{n\ge0},{\small \begin{eqnarray*}
&&f(n+d-s)+\alpha_1f(n+d-(s+1))+\cdots+\alpha_{d-s}f(n)\\
&&=A^{n+d}X+\alpha_1A^{n+d-1}X+\cdots+\alpha_{d-s}A^{n+s}X\\
&&=(A^d+\alpha_1A^{d-1}+\cdots+\alpha_{d-s}A^s)A^nX=0.
\end{eqnarray*}}
Thus, for each \sm{j},
\sm{f_j(n+d-s)+\alpha_1f_j(n+d-(s+1))+\cdots+\alpha_{d-s}f_j(n)=0.}
By Lemma \ref{Generating function},
\sm{f_j(n)=\sum\limits_{i=1}^kp_{ji}(n)\xi_i^n} and
\sm{Q(x)=1+\alpha_1x+\cdots+\alpha_{d-s}x^{d-s}=\prod\limits_{i=1}^{k}(1-\xi_ix)^{d_i}}
where \sm{p_{ji}(n)} is a polynomial in \sm{n} of degree less than
\sm{d_i}. It's easy to see that \sm{x=0} is not a solution of
\sm{Q(x)} and \sm{\sum_{i=1}^{k} d_i=d-s}.
Because{\small
\[D(x)=x^{d}Q(\frac{1}{x})=x^{d-s}+\alpha_1x^{d-s-1}+\ldots+\alpha_{d-s}
=x^{d}\prod\limits_{i=1}^{k}(1-\frac{\xi_i}{x})^{d_i}
=x^s\prod\limits_{i=1}^{k}(x-\xi_i)^{d_i},\]} \sm{\xi_i}'s are all
the distinct nonzero complex eigenvalues of \sm{A} and \sm{d_i} is
the multiplicity of \sm{\xi_i}. That completes the proof.
\end{proof}

\begin{remark}\label{re:fn}
According to Corollary \ref{eq:fn}, we may compute the general
expression of \sm{A^nX} as follows. First, compute all the complex
eigenvalues of \sm{A} and their multiplicities. Second, suppose each
\sm{f_j(n)} of \sm{f(n)} is in the form of eq.(\ref{eq:fn}) where
the coefficients of \sm{p_{ji}} are to be computed. Third, compute
\sm{f(1),\ldots, f(d)}, and obtain a set of linear equations by
comparing the coefficients of the resulted \sm{f_j(i)~ (1\le i\le
d)} to those of eq.(\ref{eq:fn}). Finally, by solving those linear
equations, we can obtain \sm{f_j(n)} and \sm{f(n)}.
\end{remark}

\begin{example} Let's consider the following loop:
{\small \[{\bf while}\quad
(x_5+x_1^2+x_1x_2-x_3^2-2x_3x_4-x_4^2>0)\quad \{X:=AX;\},\] } where
{\small \[A=\begin{bmatrix}1&-\frac{2}{5}&0&0&0\\2&\frac{1}{5}&0&0&0\\
0&0&0&2&0\\0&0&-\frac{1}{2}&-1&0\\0&0&0&0&-\frac{1}{2}\end{bmatrix}.\]}
\oomit{\[\left\{\begin{bmatrix}x_1\\x_2\\x_3\\x_4\\x_5\end{bmatrix}:=\begin{bmatrix}1&-\frac{2}{5}&0&0&0\\2&\frac{1}{5}&0&0&0\\
0&0&0&2&0\\0&0&-\frac{1}{2}&-1&0\\0&0&0&0&-\frac{1}{2}\end{bmatrix}\begin{bmatrix}x_1\\x_2\\x_3\\x_4\\x_5\end{bmatrix}\right\}.
\]} We shall show how to compute the general
expression of \sm{A^nX} by taking use of Corollary \ref{co:fn}. The
characteristic polynomial of \sm{A} is{\small
\[D(\lambda)=\lambda^5+\frac{3}{10}\lambda^4+\frac{7}{10}\lambda^3+\frac{1}{5}\lambda^2+\frac{9}{10}\lambda+\frac{1}{2}
=(\lambda+\frac{1}{2})(\lambda^2-\frac{6}{5}\lambda+1)(\lambda^2+\lambda+1).\]}
The eigenvalues of \sm{A} are {\small
\[\xi_1=\frac{3+4\ii}{5}, \xi_2=\frac{3-4\ii
}{5}, \xi_3=-\frac{1}{2}+\frac{\sqrt{3}}{2}\ii,
\xi_4=-\frac{1}{2}-\frac{\sqrt{3}}{2}\ii, \xi_5=-\frac{1}{2}.\]} Set
\sm{f(n)=A^nX=\begin{bmatrix}f_1(n)&f_2(n)&f_3(n)&f_4(n)&f_5(n)\end{bmatrix}^T}.
For any \sm{n>0}, we have {\small
\[f(n+5)+\frac{3}{10}f(n+4)+\frac{7}{10}f(n+3)+\frac{1}{5}f(n+2)+\frac{9}{10}f(n+1)+\frac{1}{2}f(n)=0.\]}
Because the multiplicities of \sm{\xi_1,\xi_2,\xi_3,\xi_4} and
\sm{\xi_5} are all $1$, by Corollary \ref{co:fn}, we may assume for
\sm{1\le j\le 5}{\small
\[f_j(n)=(\sum\limits_{i=1}^5a_{ji}x_i)\xi_1^n+(\sum\limits_{i=1}^5b_{ji}x_i)\xi_2^n
+(\sum\limits_{i=1}^5c_{ji}x_i)\xi_3^n+(\sum\limits_{i=1}^5d_{ji}x_i)\xi_4^n+(\sum\limits_{i=1}^5e_{ji}x_i)\xi_5^n.\\
\]}
Let \sm{f(1),f(2),f(3),f(4)} and \sm{f(5)} equal to
\sm{AX,A^2X,A^3X,A^4X} and \sm{A^5X} respectively, and by solving
some linear equations (see Remark \ref{re:fn}) we can obtain{\small
\begin{eqnarray*}
    &&f_1(n)=\left(\frac{2-\ii}{4}x_1+\frac{\ii}{4}x_2\right)\xi_1^n+\left(\frac{2+\ii}{4}x_1-\frac{\ii}{4}x_2\right)\xi_2^n,\\
    &&f_2(n)=\left(\frac{-5\ii}{4}x_1+\frac{2+\ii}{4}x_2\right)\xi_1^n+\left(\frac{5\ii}{4}x_1+\frac{2-\ii}{4}x_2\right)\xi_2^n,\\
    &&f_3(n)=\left((\frac{1}{2}-\frac{\sqrt{3}\ii}{6})x_3-\frac{2\sqrt{3}\ii}{3}x_4\right)\xi_3^n+\left((\frac{1}{2}+\frac{\sqrt{3}\ii}{6})x_3+\frac{2\sqrt{3}\ii}{3}x_4\right)\xi_4^n,\\
    &&f_4(n)=\left(\frac{\sqrt{3}\ii}{6}x_3+(\frac{1}{2}+\frac{\sqrt{3}\ii}{6})x_4\right)\xi_3^n+\left(-\frac{\sqrt{3}\ii}{6}x_3+(\frac{1}{2}-\frac{\sqrt{3}\ii}{6})x_4\right)\xi_4^n,\\
    &&f_5(n)=x_5\xi_5^n.
\end{eqnarray*}}
\end{example}
\subsection{Main Algorithm}\label{sec:main}

According to subsection \ref{sec:An}, the general expression of
$A^{n+m}X$ is a polynomial in $x_1, \ldots,x_N,n$ and
$\xi_1,\ldots,\xi_{q}$, the nonzero complex roots of $D(x)$. If we
substitute $A^{n+m}X$ for $X$ in $P(X)$ and denote the resulted
$P_j(X) (1\le j\le m)\in P(X)$ by $P_j(X,n)$, then $P_j(X,n)$ can be
written as
\begin{equation}\label{eq:pi}
P_j(X,n)=p_{j0}(X,n)+p_{j1}(X,n)\eta_1^n+\ldots+p_{jM}(X,n)\eta_M^n,
\end{equation}
where $\eta_k(1\le k\le M)$ is the product of some $\xi_j$'s.


To determine whether \Pone\ terminates, we have to determine for
each $j$ whether $P_j(X,n)>0$ holds for all $n$. To this end, it is
sufficient to know whether the dominant term ({\em leading term}) of
eq.(\ref{eq:pi}) is positive or not as $n\rightarrow +\infty.$ In
the following, we shall give a more detailed description of
eq.(\ref{eq:pi}) so that we can obtain the expression of the leading
term of eq.(\ref{eq:pi}).

Let \sm{\eta_k=r_ke^{\alpha_k2\pi\ii}}, where \sm{\ii=\sqrt{-1}} and
$r_k$ is the modulus of $\eta_k$. Without loss of generality, we
assume $r_1<r_2<\cdots<r_M$. For convenience, set $\eta_0=r_0=1$.
Rewrite $P_j(X,n)$ as {\small
\begin{equation*}
P_j(X,n)= p_{j0}(X,n)r_0^n+p_{j1}(X,n)e^{n\alpha_12\pi\ii}
r_1^n+\cdots+p_{jM}(X,n)e^{n\alpha_M2\pi\ii}r_M^n.
\end{equation*} }
Suppose \sm{T_j} is the common period of all the
\sm{e^{\alpha_q2\pi\ii}\ (1\le q\le M)} where \sm{\alpha_q} is a
rational number.

\begin{definition}
For each \sm{j\,(1\le j\le \sum\limits_{l=1}^{m}T_l),} if
\sm{j=\sum\limits_{l=1}^{s-1}T_l+i} and \sm{1\le i\le T_s}, then
define{\small
\[G_{j}(X,n)\triangleq P_s(X,T_sn+i-1).\]}
\end{definition}

\begin{notation}
For each \sm{j\ (1\le j\le \sum\limits_{l=1}^{m}T_l)}, expand
\sm{G_j(X,n)}, collect the result with respect to (w.r.t.)
\sm{n^lr_k^n}, and let \sm{C_{jkl}(X,n)} denote the coefficient of
the term \sm{n^lr_k^n}.
\end{notation}

Then \sm{G_j(X,n)} can be written as {\small
\begin{eqnarray}\label{eq:cj}
C_{j10}(X,n)r_1^n+C_{j11}(X,n)nr_1^n+\cdots+C_{j1d_1}(X,n)n^{d_1}r_1^n+\cdots\\
\nonumber
+C_{jM0}(X,n)r_{M}^n+C_{jM1}(X,n)nr_{M}^n+\cdots+C_{jMd_{M}}(X,n)n^{d_{M}}
r_{M}^n,\end{eqnarray} } where \sm{d_l (1\le l\le M)} is the
greatest degree of \sm{n} in \sm{G_j(X,n)} w.r.t. \sm{r_l}.

It can be deduced that if $r_i<r_j$, the order of $n^{l_1}r_i^n$ is
less than the order of $n^{l_2}r_j^n$ for any $l_1$ and $l_2$ when
$n$ goes to infinity. Similarly, if $l_1<l_2$, the order of
$n^{l_1}r_i^n$ is less than the order of $n^{l_2}r_i^n$. So, it is
natural to introduce an ordering on the terms $n^{l}r_j^n$ as
follows.

\begin{definition}
We define $n^{l_1}r_i^n\prec n^{l_2}r_j^n$ if $r_i<r_j$ or $r_i=r_j$
and $l_1<l_2$. A term $C_{jkl}n^lr_j^n$ in eq.{\rm (\ref{eq:cj})} is
said to be the {\em leading term} and $C_{jkl}$ the {\em leading
coefficient} if $n^lr_j^n$ occurring in $C_{jkl}$ is the largest one
under that ordering $\prec$.
\end{definition}

Suppose \sm{G_j(X,n)=P_{s}(X,T_{s}n+t)} for some \sm{s} and \sm{t}.
For those \sm{e^{\alpha_q2\pi\ii}}'s where \sm{\alpha_q}'s are
rational numbers,
\sm{e^{(T_sn+t)\alpha_q2\pi\ii}=e^{t\alpha_q2\pi\ii}} because
\sm{T_s} is the common period. Because there may be some
\sm{e^{(T_sn+t)\alpha_q2\pi\ii}}'s with irrational \sm{\alpha_q}'s,
each \sm{C_{jkl}(X,n)} can be divided into three parts,{\small
\[C_{jkl}(X,n)=C_{jkl0}(X)+C_{jkl1}(X)+C_{jkl2}(X,n),\]} where
\sm{C_{jkl0}(X)} does not contain any
\sm{e^{(T_sn+t)\alpha_q2\pi\ii}}, \sm{C_{jkl1}(X)} contains
\sm{e^{(T_sn+t)\alpha_q2\pi\ii}} with rational \sm{\alpha_q} and
\sm{C_{jkl2}(X,n)} contains those \sm{e^{(T_sn+t)\alpha_q2\pi\ii}}
with irrational \sm{\alpha_q}\footnote{Later it will be proved that
\sm{C_{jkl0}(X), C_{jkl1}(X)}
and \sm{C_{jkl2}(X,n)} are reals for any \sm{n}.}. 
Further, \sm{C_{jkl2}(X,n)} can be written as {\small
\begin{eqnarray*}C_{jkl2}(X,n)=C_{jkl2}(X,\sin((nT_s+t)\alpha_{k1}2\pi),\cos((nT_s+t)\alpha_{k1}2\pi),\ldots,\\
\sin((nT_s+t)\alpha_{ks_k}2\pi),\cos((nT_s+t)\alpha_{ks_k}2\pi)),\end{eqnarray*}
} where \sm{\{\alpha_{k1}2\pi,\ldots,\alpha_{ks_k}2\pi\}} is a
maximum {\em rationally independent} group\footnote{``Rationally
independent" will be described later.}.

\begin{example} We continue to use the loop in Example 1 to illustrate the above
concepts and notations.

Because \sm{|\xi_1|=|\xi_2|=1}, let \sm{\xi_1=e^{\alpha_12\pi\ii}}
and \sm{\xi_2=e^{-\alpha_12\pi \ii}}, where \sm{\alpha_12\pi} is the
argument of \sm{\xi_1}. It's not difficult to check that
\sm{\alpha_1} is an irrational number.\footnote{A general algorithm
for checking whether an argument is a rational multiple of $\pi$
will be stated in detail in subsection 3.4.} For the sake of
clarity, in the following we firstly reduce the expressions of
\sm{f_1(n),f_2(n), f_3(n),f_4(n)} and \sm{f_5(n)}, and then
substitute them in the loop guard. Let \sm{\alpha_2=\frac{1}{3}},
then \sm{\xi_3=e^{\alpha_22\pi\ii}},
\sm{\xi_4=e^{-\alpha_22\pi\ii}}, and
 \sm{f_1(n), f_2(n), f_3(n), f_4(n)} and \sm{f_5(n)} can be
rewritten as {\small
\begin{eqnarray*}
     &&f_1(n)=x_1\cos(n\alpha_12\pi)+\frac{x_1-x_2}{2}\sin(n\alpha_12\pi),\\
     &&f_2(n)=x_2\cos(n\alpha_12\pi)+\frac{5x_1-x_2}{2}\sin(n\alpha_12\pi),\\
     &&f_3(n)=x_3\cos(n\alpha_22\pi)+\frac{\sqrt{3}x_3+4\sqrt{3}x_4}{3}\sin(n\alpha_22\pi),\\
     &&f_4(n)=x_4\cos(n\alpha_22\pi)-\frac{\sqrt{3}x_3+\sqrt{3}x_4}{3}\sin(n\alpha_22\pi),\\
     &&f_5(n)=x_5(-\frac{1}{2})^n.
\end{eqnarray*}}
Substituting \sm{f_1(n),f_2(n),f_3(n),f_4(n)} and \sm{f_5(n)} for
\sm{x_1,x_2,x_3,x_4} and \sm{x_5} respectively in the loop guard, we
get that the resulted loop guard is
\[L_1r_1^n+(L_{21}+L_{22}+L_{23})r_2^n,\]
where \sm{r_1=\frac{1}{2},r_2=1}, and {\small
\begin{eqnarray*}
     &&L_1=(-1)^nx_5,\\
     &&L_{21}=-\frac{x_3^2+2x_3x_4+4x_4^2}{2}+\frac{5x_1^2+x_2^2-2x_1x_2}{4},\\
     &&L_{22}=\frac{2x_4^2-2x_3x_4-x_3^2}{2}\cos(n2\alpha_22\pi)-(\sqrt{3}x_3x_4+\sqrt{3}x_4^2)\sin(n2\alpha_22\pi),\\
     &&L_{23}=-\frac{x_1^2+x_2^2-6x_1x_2}{4}\cos(n2\alpha_12\pi)+\frac{7x_1^2-x_2^2-2x_1x_2}{4}\sin(n2\alpha_12\pi).
\end{eqnarray*}}
\oomit{{\small
\[((-1)^nx_5)(\frac{1}{2})^n+(-\frac{x_3^2+2x_3x_4+4x_4^2}{2}+\frac{5x_1^2+x_2^2-2x_1x_2}{4}+
\frac{2x_4^2-2x_3x_4-x_3^2}{2}\cos(n2\alpha_22\pi)-\]\[(\sqrt{3}x_3x_4+\sqrt{3}x_4^2)\sin(n2\alpha_22\pi)
-\frac{x_1^2+x_2^2-6x_1x_2}{4}\cos(n2\alpha_12\pi)+\frac{7x_1^2-x_2^2-2x_1x_2}{4}\sin(n2\alpha_12\pi))1^n.\]}}
Since \sm{\alpha_2=\frac{1}{3}}, the period of
\sm{\xi_3^2=e^{2\alpha_22\pi}=\cos(2\alpha_22\pi)+\sin(2\alpha_22\pi)\ii}
is 3. Then we compute \oomit{{\small \begin{eqnarray*}
G_1(X,n)=((-1)^{3n}x_5)(\frac{1}{2})^{3n}+(-\frac{x_3^2+2x_3x_4+4x_4^2}{2}+\frac{5x_1^2+x_2^2-2x_1x_2}{4}+
\frac{2x_4^2-2x_3x_4-x_3^2}{2}-\\\frac{x_1^2+x_2^2-6x_1x_2}{4}\cos(3n2\alpha_12\pi)+\frac{7x_1^2-x_2^2-2x_1x_2}{4}\sin(3n2\alpha_12\pi))1^{3n},
\end{eqnarray*}
\begin{eqnarray*}G_2(X,n)=((-1)^{3n+1}x_5)(\frac{1}{2})^{3n+1}+(-\frac{x_3^2+2x_3x_4+4x_4^2}{2}+\frac{5x_1^2+x_2^2-2x_1x_2}{4}+
\frac{-2x_4^2+2x_3x_4+x_3^2}{4}+\\(\frac{3}{2}x_3x_4+\frac{3}{2}x_4^2)-\frac{x_1^2+x_2^2-6x_1x_2}{4}\cos((3n+1)2\alpha_12\pi)+\frac{7x_1^2-x_2^2-2x_1x_2}{4}\sin((3n+1)2\alpha_12\pi))1^{3n+1},
\end{eqnarray*}
\begin{eqnarray*}
G_3(X,n)=((-1)^{3n+2}x_5)(\frac{1}{2})^{3n+2}+(-\frac{x_3^2+2x_3x_4+4x_4^2}{2}+\frac{5x_1^2+x_2^2-2x_1x_2}{4}+
\frac{-2x_4^2+2x_3x_4+x_3^2}{4}-\\(\frac{3}{2}x_3x_4+\frac{3}{2}x_4^2)-
\frac{x_1^2+x_2^2-6x_1x_2}{4}\cos((3n+2)2\alpha_12\pi)+\frac{7x_1^2-x_2^2-2x_1x_2}{4}\sin((3n+2)2\alpha_12\pi))1^{3n+2},
\end{eqnarray*}}}
{\small
\begin{eqnarray*}
&& G_1(X,n)=((-1)^{3n}x_5)r_1^{3n}+C_{120}r_2^{3n},\\
&& G_2(X,n)=((-1)^{3n+1}x_5)r_1^{3n+1}+C_{220}r_2^{3n+1},\\
&& G_3(X,n)=((-1)^{3n+2}x_5)r_1^{3n+2}+C_{320}r_2^{3n+2}.
\end{eqnarray*}
} Take \sm{G_1(X,n)} as an example.
{\small
\begin{eqnarray*}
   &&C_{120}=C_{1200}+C_{1201}+C_{1202},\\
   &&C_{1200}=-\frac{x_3^2+2x_3x_4+4x_4^2}{2}+\frac{5x_1^2+x_2^2-2x_1x_2}{4},\\
   &&C_{1201}=\frac{2x_4^2-2x_3x_4-x_3^2}{2},\\
   &&C_{1202}=-\frac{x_1^2+x_2^2-6x_1x_2}{4}\cos(3n2\alpha_12\pi)+\frac{7x_1^2-x_2^2-2x_1x_2}{4}\sin(3n2\alpha_12\pi).
\end{eqnarray*}}
\end{example}

\begin{notation}\label{succ} 
We denote by \sm{C_{jkl}(X,n)\succ 0} (and call \sm{C_{jkl}(X,n)}
``positive") if{\small
\[\min\{C_{jkl0}(X)+C_{jkl1}(X)+C_{jkl2}(X,y_{11},y_{12},\ldots,y_{s_k1},y_{s_k2})\}>0\]
} subject to \sm{\{y_{11}^2+y_{12}^2=1,\cdots,
y_{s_k1}^2+y_{s_k2}^2=1\}.}
\end{notation}

\begin{remark}\label{succremark} It is not difficult to see \sm{C_{jkl}(X,n)\succ 0}
iff {\small
\[\begin{array}{r}\smallskip
\forall (y_{11},y_{12},\ldots,y_{s_k1},y_{s_k2})\
((y_{11}^2+y_{12}^2=1)\wedge\cdots\wedge(y_{s_k1}^2+y_{s_k2}^2=1)\rightarrow\\
C_{jkl0}(X)+C_{jkl1}(X)+C_{jkl2}(X,y_{11},y_{12},\ldots,y_{s_k1},y_{s_k2})>0)
\end{array}\] }
because
\sm{\{(y_{11},y_{12},\ldots,y_{s_k1},y_{s_k2}):y_{i1}^2+y_{i2}^2=1,i=1,...,s_k\}}
is a bounded closed set. If \sm{\succ} and \sm{>} are replaced with
\sm{\succeq} and \sm{\ge} respectively in the notation, we get the
notation of \sm{C_{jkl}(X,n)\succeq 0} (``nonnegative").
\end{remark}

Roughly speaking, for any \sm{G_j(X,n)}, if its leading coefficient
\sm{C_{jkl}(X,n)\succ 0}, there exists an integer \sm{N_1} such that
for all \sm{n>N_1}, \sm{G_j(X,n)> 0}. If all the leading
coefficients of all the \sm{G_j(X,n)}'s are ``positive", there
exists \sm{N'} such that for all \sm{n>N'}, all the \sm{G_j(X,n)}'s
are positive. Therefore, \Pone\ is nonterminating with input
\sm{X':=A^{N'}X}.

On the other hand, if \Pone\ is nonterminating, does there exist an
input \sm{X} such that the leading coefficients of all the
\sm{G_j(X,n)}'s are ``positive"? We do not know the answer yet.
However, with an assumption described below, the answer is yes.

{\tt Assumption for the main algorithm:} For any
\sm{X\in\mathbb{R}^n} and any \sm{C_{jkl}(X,n)}, \sm{C_{jkl2}(X,n)}
being not identically zero implies {\small
\[\min(C_{jkl0}(X)+C_{jkl1}(X)+C_{jkl2}(X,y_{11},y_{12},\ldots,y_{s_k1},y_{s_k2}))\ne0\]
}
subject to \sm{y_{i1}^2+y_{i2}^2=1\ (i=1,\ldots, s_k)}.

It is
not difficult to see that the assumption is equivalent to the
following formula:{\small
\[\begin{array}{rl}
\left[\forall X\forall Y\right. & (C_{jkl2}(X,n)\equiv
0\vee\\
& \left.\bigwedge_{1\le i\le s_k}y_{i1}^2+y_{i2}^2=1\rightarrow
C_{jkl0}(X)+C_{jkl1}(X)+C_{jkl2}(X,Y)>0)\right]~\bigvee\\
\left[\forall X\exists Y\right. & (C_{jkl2}(X,n)\equiv
0\vee\\
& \left.\bigwedge_{1\le i\le s_k}y_{i1}^2+y_{i2}^2=1\rightarrow
C_{jkl0}(X)+C_{jkl1}(X)+C_{jkl2}(X,Y)<0)\right].\end{array}\]}
Because \sm{C_{jkl2}(X,n)} can be written as \sm{\sum\limits_{i\in
I}f_{i1}(X)\sin(n\alpha_i 2\pi)+f_{i2}(X)\cos(n\alpha_i2\pi)}, where
\sm{I} is an index set, \sm{C_{jkl2}(X,n)\equiv 0} is equivalent to
\sm{\bigwedge_{i\in I}f_{i1}(X)=f_{i2}(X)=0}. Thus, the assumption
can be checked with real quantifier elimination techniques.

\begin{example}
For those \sm{C_{ijk}}'s in Example 2, let's check whether they
satisfy the assumption. For clarity, we present a clear proof here
rather than make use of any tool for real quantifier elimination.
Take \sm{G_1(X,n)} for example. It's clear that
\sm{C_{1202}(X,n)\equiv 0} if and only if \sm{x_1=x_2=0}. It is not
difficult to compute {\small
\[D=\inf\limits_{n\ge1}C_{1202}(X,n)=-\sqrt{\left(\frac{-x_1^2-x_2^2+6x_1x_2}{4}\right)^2+\left(\frac{7x_1^2-x_2^2-2x_1x_2}{4}\right)^2}.\]}
If \sm{x_1=x_2=0} does not hold,
\sm{\frac{5x_1^2+x_2^2-2x_1x_2}{4}+D<0} because {\small
\[(\frac{5x_1^2+x_2^2-2x_1x_2}{4})^2-D^2=\frac{-1}{16}(5x_1^2+x_2^2-2x_1x_2)^2<0.\]}
Let{\small
\[M=\min\limits_{y_{11}^2+y_{12}^2=1}(C_{1200}(X)+C_{1201}(X)+C_{1202}(X,y_{11},y_{12})),\]}
then {\small
\[M=-(x_3+x_4)^2+\frac{5x_1^2+x_2^2-2x_1x_2}{4}+D<0\]
}if \sm{x_1=x_2=0} does not hold. Consequently, \sm{G_1(X,n)}
satisfies the assumption. Similarly it can be proved \sm{G_2(X,n)}
and \sm{G_3(X,n)} satisfy the assumption too. Thus the loop in
Example 1 satisfies the assumption.
\end{example}

We shall show in next section how hard it is to deal with the case
that the assumption does not hold. Now, we are ready to describe our
main algorithm. For the sake of brevity, the algorithm is described
as a nondeterministic algorithm. The basic idea is to guess a
leading term for each \sm{G_j(X,n)} first. Then, setting its
coefficient be ``positive" and the coefficients of the terms with
higher order be ``nonnegative", we can get a semi-algebraic system
(SAS). If one of our guess is satisfiable, {\it i.e.}, one of the
SASs has solutions, \Pone\ is nonterminating. Otherwise, it is
terminating.

\medskip
\noindent\textbf{Algorithm} ~~ \texttt{Termination}
\begin{description}
\item[Step 0] Compute the general expression of \sm{A^{n+m}X}.
\item[Step 1] Substitute \sm{A^{n+m}X} for \sm{X} in \sm{P(X)}, and compute all
\sm{G_j(X,n)} (finite many, say, \sm{j=1,...,L}).  
\item[Step 2] Guess a leading term for each \sm{G_j(X,n)}, say
\sm{C_{jk_jl_j}n^{l_j}r_{k_j}^n}.
\item[Step 3] Construct a semi-algebraic system \sm{S} as follows.
{\small
\[\begin{array}{l} S_j=C_{jk_jl_j}\succ 0 \wedge \bigwedge_{(k>k_j) \vee (k=k_j\wedge l>l_j)} C_{jkl}(X,n)\succeq
0,~~ S=\bigwedge_{j=1}^{L} S_j.
\end{array}\]
 }
\item[Step 4] If one of these systems is satisfiable, return
"nonterminating". Otherwise return "terminating".
\end{description}

\begin{remark}
If {\tt the assumption for the main algorithm} does not hold, then
{\tt Termination} is incomplete. That is, if it returns
``nonterminating", the loop is nonterminating. Otherwise, it tells
nothing.
\end{remark}

\begin{example} For the loop in Example 1, we have computed the
\sm{G_i(X,n)}'s in Example 2 and verified that it satisfies the
assumption of the main algorithm in Example 3. We shall finish its
termination decision in this example, following the steps in {\tt
Termination}.

By Steps 2 and 3 of {\tt Termination}, we should guess leading terms
and construct SASs accordingly. To be concrete, let's take as an
example one certain guess and suppose we have the following
semi-algebraic system {\small
\begin{equation*}
\left\{\begin{array}{l}C_{120}(X,n)\succ 0,\\
C_{220}(X,n)\succ 0,\\
C_{320}(X,n)\succ 0.
\end{array}
\right.
\end{equation*}}
According to Notation \ref{succ} and Remark \ref{succremark}, the
above inequalities are equivalent to \sm{(\forall y_{11},y_{12})
~y_{11}^2+y_{12}^2=1\rightarrow} {\small
\begin{equation*}
\left\{\begin{array}{l}-(x_3+x_4)^2+\frac{5x_1^2+x_2^2-2x_1x_2}{4}-\frac{x_1^2+x_2^2-6x_1x_2}{4}y_{11}+\frac{7x_1^2-x_2^2-2x_1x_2}{4}y_{12}>0,\\
-(x_4-\frac{x_3}{2})^2+\frac{5x_1^2+x_2^2-2x_1x_2}{4}-\frac{x_1^2+x_2^2-6x_1x_2}{4}y_{11}+\frac{7x_1^2-x_2^2-2x_1x_2}{4}y_{12}>0,\\
-(x_4+\frac{x_3}{2})^2+\frac{5x_1^2+x_2^2-2x_1x_2}{4}-\frac{x_1^2+x_2^2-6x_1x_2}{4}y_{11}+\frac{7x_1^2-x_2^2-2x_1x_2}{4}y_{12}>0.
\end{array}
\right.
\end{equation*}}
In Example 3, we have shown that
\sm{\frac{5x_1^2+x_2^2-2x_1x_2}{4}+D\le 0}. Thus, the above
predicate formula does not hold. In fact, none of the formulas
obtained in Step 3 holds. Thus, the loop in Example 1 is
terminating.
\end{example}

\begin{remark}
It is well known that real quantifier elimination is decidable from
Tarski's work \cite{ta51}. Therefore, the semi-algebraic systems in
Step 3 can be solved. For the tools for solving semi-algebraic
systems, please be referred to \cite{co75,CH,Dol,xia07}.
\end{remark}

\begin{remark}
There are some techniques to decrease the amount of computation of
the algorithm \texttt{Termination}. For example, we can use Lemma
\ref{Braverman} when guessing leading terms for each \sm{G_j(X,n)}.
\end{remark}

\begin{lemma}{\rm \cite{Mark06}}\label{Braverman}
Let \sm{\xi_1,\xi_2,\ldots,\xi_m\in \mathbb{C}} be a collection of
distinct complex numbers such that \sm{|\xi_i|=1} and \sm{\xi_i\ne
1} for all \sm{i}. Let \sm{\alpha_1,\alpha_2,\ldots,\alpha_m} be any
complex numbers and \sm{z_n=\alpha_1\xi_1^n+\ldots+\alpha_m\xi_m^n.}
Then one of the following is true:
\begin{enumerate}
\item the real part \sm{Re(z_n)=0} for all \sm{n}; or
\item there is \sm{c<0} such that \sm{Re(z_n)<c} for infinitely many n's.
\end{enumerate}
\end{lemma}

According to Lemma \ref{Braverman}, if \sm{C_{jkl0}=0} and
\sm{C_{jkl}(X,n)} is not identically zero w.r.t. \sm{n}, then
\sm{C_{jkl}(X,n)} can not be always nonnegative. According to the
former discussion {\small \[\{G_{\sum
\limits_{k=1}^{j-1}T_k+1}(X,n),\ldots,G_{\sum\limits_{k=1}^{j-1}T_k+T_j}(X,n)\}\]
} are obtained from \sm{P_j(X,n)}. Then the leading term with the
greatest order among all the leading terms of the above
set\oomit{$\{G_{\sum
\limits_{k=1}^{j-1}T_k+1}(X,n),\ldots,G_{\sum\limits_{k=1}^{j-1}T_k+T_j}(X,n)\}$}
should not be of the form $C_{jk_jl_j}n^{l_j}r_{k_j}^n$ where
$C_{jk_jl_j0}=0$. Thus, this should be avoided when guessing leading
terms for each $G_j$.

\oomit{\begin{remark}\label{re:mini} At Step 1 of the main
algorithm, we may compute \sm{\eta_k(1\le k\le M)} which are the
products of some \sm{\xi_j}'s after substituting the general
expression of \sm{A^{n+m}X} for \sm{X} in \sm{P(X)}. In order to
describe \sm{\eta_k}, we need the minimal polynomials of
\sm{\eta_k}. In \cite{Str} Strzebonski has given an algorithm to do
this numerically. Actually, we have a general symbolic algorithm for
computing the minimal polynomials of \sm{\alpha+\beta, \alpha-\beta,
\alpha\cdot\beta} and \sm{\frac{\alpha}{\beta}} for given algebraic
numbers \sm{\alpha} and \sm{\beta}.
\end{remark}

\begin{remark}
At Step 1 of the main algorithm, \sm{T_s} is necessary for defining
\sm{G_j(X,n)}. Thus for a given \sm{\eta_k} we must check whether
its argument is a rational multiple of \sm{\pi} and if it is, we
need to know the period of $\frac{\eta_k}{|\eta_k|}$. This can be
done through exact computation.
\end{remark}}

In the following subsections, we shall explain the details of the
main steps of \texttt{Termination} and prove its correctness.

\subsection{Compute the Minimal Polynomials of $\alpha+\beta, \alpha-\beta,
 \alpha\cdot\beta$ and $\frac{\alpha}{\beta}$}
At Step 1 of the main algorithm, we may compute \sm{\eta_k(1\le k\le
M)} which are the products of some \sm{\xi_j}'s after substituting
the general expression of \sm{A^{n+m}X} for \sm{X} in \sm{P(X)}. In
order to describe \sm{\eta_k}, we need the minimal polynomials of
\sm{\eta_k}. In this subsection a method is presented to solve a
more general problem.

In \cite{Str} Strzebonski gave an algorithm to compute
$\{\alpha+\beta, \alpha-\beta, \alpha\cdot\beta,
\frac{\alpha}{\beta}\}$ numerically where $\alpha$ and $\beta$ are
given algebraic numbers. Here we want to present another more
intuitive method based on symbolic computation.

In the following let $\alpha\diamond\beta$ denote one of
$\{\alpha+\beta, \alpha-\beta, \alpha\cdot\beta,
\frac{\alpha}{\beta}\}$. Without loss of generality let's assume
that the minimal polynomial of $\alpha$ is $f_1(x)$ whose degree is
$d_1$ and the minimal polynomial of $\beta$ is $f_2(x)$ whose degree
is $d_2$. We can bound $\alpha$ and $\beta$ in $W_1$ and $W_2$,
respectively, where $W_1,W_2$ are ``boxes", by isolating the complex
zeros of $f_1(x)$ and $f_2(x)$. Since the degree of $\alpha$ is
$d_1$ and the degree of $\beta$ is $d_2$, the degree of
$\alpha\diamond\beta$ is at most $d_1\cdot d_2$. Then there must
exist $x_1,\ldots x_{d_1\cdot d_2+1}$ in $\mathbb{Z}$ such that
\[x_1+x_2\alpha\diamond\beta+\ldots+x_{d_1\cdot
d_2+1}(\alpha\diamond\beta)^{d_1\cdot d_2}=0.\] Thus we can design
an algorithm to enumerate all the $(x_1,\ldots,x_{d_1\cdot
d_2+1})\in \mathbb{Z}^{d_1\cdot d_2+1}$ and check whether it is a
solution. Since there exists one solution this algorithm must
terminate and output a solution. Assume that its output is
$a_0,a_1,\ldots,a_{d_1\cdot d_2}$ and
$f(x)=a_0+a_1x+\ldots+a_{d_1\cdot d_2}x^{d_1\cdot d_2}$. Because
$f(\alpha\diamond\beta)=0$, the minimal polynomial of
$\alpha\diamond\beta$ is an irreducible factor of $f(x)$. Factor
$f(x)$ in $\mathbb{Q}$. Without loss of generality, we assume
$f(x)=g_1(x)^{m_1}\cdot g_2(x)^{m_2}\ldots g_d(x)^{m_d}.$ We can
check whether $g_j(x)(1\le j\le d)$ is the minimal polynomial of
$\alpha\diamond\beta$ by solving the following semi-algebraic system
(SAS): \[\{g_j(x\diamond y)=0, f_1(x)=0, f_2(y)=0, x\in W_1, y\in
W_2\}.\] If it is satisfiable, $g_j(x)$ is the minimal polynomial of
$\alpha\diamond\beta$; otherwise it is not. Thus the minimal
polynomial of $\alpha\diamond\beta$ can be obtained.

\subsection{Check Whether the Argument of $\alpha$ Is a Rational
Multiple of $\pi$} At Step 1 of the main algorithm, \sm{T_s} is
necessary for defining \sm{G_j(X,n)}. Thus for a given \sm{\eta_k}
we have to check whether its argument is a rational multiple of
\sm{\pi} and if it is, we need to know the period of
$\frac{\eta_k}{|\eta_k|}$. This subsection aims at this problem.

Suppose the minimal polynomial of \sm{\alpha} is \sm{p(x)} whose
degree is \sm{d}. Without loss of generality suppose
\sm{\alpha=re^{\beta 2\pi\ii}}. We can bound $\alpha$ in $W$ by
isolating all the complex roots of $p(x)$. Since the degree of
$\alpha$ is $d$, the degree of $\overline{\alpha}$ must be $d$. Then
the degree of $\alpha\cdot\overline{\alpha}=r^2$ is at most $d^2$.
Thus the degree of $r$ is at most $2d^2$. The degree of
$\alpha^{-1}$ is at most $2d^2$ because the degree of $\alpha^{-1}$
is the same as the degree of $\alpha$. Since the degree of $\alpha$
is $d$ the degree of $\alpha\cdot r^{-1}=e^{i\beta 2\pi}$ is at most
$2d^3$.

If $\beta$ is a rational number, $\alpha$ must be a unit root and
its minimal polynomial must be a cyclotomic polynomial. As a result
if $\beta$ is a rational number the minimal polynomial of $\alpha$
must be a cyclotomic polynomial whose degree is less than or equal
to $2n^3$. All the cyclotomic polynomials can be computed explicitly
according to the theory of cyclotomic field. Thus let $CP_j(x)$
denote the cyclotomic polynomial whose degree is $j$. Then, $\beta$
is a rational number if and only if the following is satisfiable
\[\exists r((r\ne
0)\bigwedge(\bigvee\limits_{j=1}^{2d^3}CP_j(x/r)=0)\bigwedge
(p(x)=0)\bigwedge (x\in W)).\] Because
$CP_j(x/r)=0\Longleftrightarrow r^jCP_j(x/r)=0$, the above
quantifier formula is decidable. If it's satisfiable the minimal
polynomial of $\alpha$ can be computed by checking whether \[\exists
r((r\ne 0)\bigwedge(CP_j(x/r)=0)\bigwedge (p(x)=0)\bigwedge (x\in
W))\] is satisfiable one by one.

\subsection{Check Rational Independence}
\oomit{For the sub-expression,$C_{jkl2}(X,n)$, of $C_{jkl}(X,n)$ its
maximum {\em rationally independent} group is necessary and in fact
the representation expression of any other $e^{\gamma2\pi}$ in this
group of elements is necessary too. These are the intention of this
section.}

Given a set of algebraic numbers,
\sm{\alpha_1=e^{\beta_12\pi\ii},\ldots,
\alpha_d=e^{\beta_d2\pi\ii}}, where \sm{\beta_1,\ldots,\beta_d} are
irrational numbers. In this subsection we present a method to check
whether \sm{\beta_1,\ldots,\beta_d} are {\em rationally
independent}.
\begin{definition}Irrational numbers \sm{\beta_1,\ldots,\beta_d}
are {\em rationally independent} if there does not exist rational
numbers \sm{a_1,\ldots,a_d} such that
\sm{\sum\limits_{j=1}^{d}a_j\beta_j\in \mathbb{Q}}.
\end{definition}

Obviously, \sm{\beta_1,\ldots,\beta_d} are rationally independent if
and only if \sm{1,\beta_1,\ldots,\beta_d} are linearly independent
in \sm{\mathbb{Q}}. It can be deduced that
\sm{\beta_1,\ldots,\beta_d} are rationally independent if and only
if \sm{\forall (b_1,\ldots,b_d)\in \mathbb{Z}^d,
\sum\limits_{j=1}^{d}b_j\beta_j\notin \mathbb{Z}}.

\begin{lemma}{\rm \cite{Baker}}\label{baker}
Let \sm{\lambda_1,\ldots,\lambda_m} with \sm{m\ge 2} be linearly
dependent logarithms of algebraic numbers. Define
\sm{\alpha_j=e^{\lambda_j}(1\le j\le m)}. For \sm{1\le j\le m},
let \sm{\log{A_j}\ge1} be an upper bound for
\sm{\max\{h(\alpha_j),\frac{|\lambda_j|}{D}\}} where \sm{D} is the
degree of the number field
\sm{K=\mathbb{Q}(\alpha_1,\ldots,\alpha_m)} over \sm{\mathbb{Q}}
and \sm{h(\alpha)} denotes the absolute logarithmic height of
\sm{\alpha}. Then there exist rational integers
\sm{n_1,\ldots,n_m}, not all of which are zero, such that
\sm{n_1\lambda_1+\ldots+n_m\lambda_m=0} and
\sm{|n_k|<(11(m-1)D^3)^{m-1}\frac{(\log{A_1})\ldots(\log{A_m})}{\log{A_k}}}
for \sm{1\le k\le m}.
\end{lemma}
\begin{remark}
Baker is the first one to use his transcendence arguments to
establish such an estimate. However, the description here follows
Lemma 7.19 in \cite{wal}.
\end{remark}

Sequence \sm{\{\beta_1,\ldots,\beta_d,1\}} are linearly independent
in \sm{\mathbb{Q}} iff \sm{\{\beta_12\pi
\ii,\ldots,\beta_d2\pi\ii,2\pi\ii\}} are linearly independent in
\sm{\mathbb{Q}}. If
\sm{n_1\beta_12\pi\ii+\ldots+n_d\beta_d2\pi\ii+n_{d+1}2\pi\ii=0,}
then\\ \sm{e^{n_1\beta_12\pi\ii}\cdots e^{n_d\beta_d2\pi\ii}=1.}
That is \sm{\alpha_1^{n_1}\cdots\alpha_d^{n_d}=1}. According to
Lemma \ref{baker} we can decide whether
\sm{\{\beta_1,\ldots,\beta_d\}} are rationally independent by
enumerating \sm{n_k} from  {\small \[
  \lfloor-(11dD^3)^{d}\frac{(\log{A_1})\cdots(\log{A_{d+1}})}{\log{A_k}}\rfloor,\ldots,
  \lceil(11(d)D^3)^{d}\frac{(\log{A_1})\cdots(\log{A_{d+1}})}{\log{A_k}}\rceil\]
  }
for \sm{k=1,\ldots,d} and checking whether
\sm{\alpha_1^{n_1}\cdots\alpha_d^{n_d}=1}. For any given
\sm{(n_1,\ldots,n_d)}, whether
\sm{\alpha_1^{n_1}\cdots\alpha_d^{n_d}=1} can be determined by
checking whether the following SAS has solutions.
{\small\[\begin{array}{c}\{x_1^{n_1}\cdots x_d^{n_d}=1,~
q_j(x_j)=0,~x_j\in W_j,~ j=1,...,d\},\end{array}\]} where \sm{q_j}
is the minimal polynomial of \sm{\alpha_j} and \sm{W_j} contains
only one complex root of $q_j$ for \sm{j=1,\ldots,d}.

\oomit{If there exists \sm{(n_1,\ldots,n_d)} such that
\sm{\alpha_1^{n_1}\ldots\alpha_d^{n_d}=1}, then they are rationally
dependent; otherwise they are rationally independent. If they are
rationally dependent, the representation of one by the others can be
obtained.}

\subsection{Compute the Infimum of $C_{jkl2}$}

To prove the correctness of our main algorithm, we need some further
results.

First, let's introduce a lemma in ergodic theory. Let \sm{S} be the
unit circumference. Usually any point on \sm{S}, say \sm{(a,b)}, is
denoted as a complex number \sm{a+b\ii}. Define \sm{\pi:
\RR^m\rightarrow S^m} as {\small \[(x_1,\ldots,x_m)\rightarrow
(e^{x_12\pi\ii},\ldots, e^{x_m2\pi\ii}).\]} Define m-torus
\sm{T^m=\{(e^{x_12\pi\ii},\ldots, e^{x_m2\pi\ii})|\ x_j\in
\mathbb{R}\}} and \sm{L_{\pi(\alpha)}:T^m\rightarrow T^m} as {\small
\[(e^{y_12\pi\ii},\ldots, e^{y_m2\pi\ii})\rightarrow
(e^{(y_1+\alpha_1)2\pi\ii},\ldots, e^{(y_m+\alpha_m)2\pi\ii}).\]}
\begin{lemma}{\rm \cite{Mane}}\label{ergodic} If \sm{\alpha\in \mathbb{R}^m}, the translation
\sm{L_{\pi(\alpha)}(X)} is ergodic iff for all \sm{K\in \ZZ^m}, \sm{
(K,\alpha)\notin \mathbb{Z}} where \sm{(K,\alpha)} stands for the
inner product of \sm{K} and \sm{\alpha}.
\end{lemma}
According to Lemma \ref{ergodic} we know that if $\alpha_1$ is an
irrational number, the closure of $\{e^{n\alpha_12\pi\ii}\}_{n\ge1}$
is the unit circumference. Also, if \sm{\alpha_1,\ldots,\alpha_m}
are rationally independent, \sm{L_{\pi(\alpha)}(X)} is ergodic. Thus
the closure of \sm{\{L_{\pi(\alpha)}^n(0)\}_{n\geq 1}} is \sm{T^m}.

\begin{lemma}\label{infimumL}
If \sm{\alpha_{k1},\ldots,\alpha_{ks_k}} are rationally independent
and \sm{C_{jkl2}} is of the form {\small
\[C_{jkl2}(X,\sin(n\alpha_{k1}2\pi),\cos(n\alpha_{k1}2\pi),\ldots,\sin(n\alpha_{ks_k}2\pi),\cos(n\alpha_{ks_k}2\pi))\]}
for a fixed \sm{X}, then{\small
\[\inf\limits_{n\ge1}\{C_{jkl2}(X,\sin(n\alpha_{k1}2\pi),\cos(n\alpha_{k1}2\pi),\ldots,\sin(n\alpha_{ks_k}2\pi),\cos(n\alpha_{ks_k}2\pi))\}\]}
is equal to
\sm{\min\{C_{jkl2}(X,x_{1},y_{1},\ldots,x_{s_k},y_{s_k})\}} subject
to \sm{\{x_i^2+y_i^2=1,1\le i\le s_k\}}.
\end{lemma}
\begin{proof} According to Lemma \ref{ergodic} for any $
(x_1,y_1,\ldots,x_{s_k},y_{s_k})$ there exists a subsequence, say
$\{n_i\}_{i\ge1}$, such that {\small
\begin{eqnarray*} && \lim\limits_{i\rightarrow
+\infty}(\sin(n_i\alpha_{k1}2\pi),\cos(n_i\alpha_{k1}2\pi),\ldots,
\sin(n_i\alpha_{ks_k}2\pi),\cos(n_i\alpha_{ks_k}2\pi))\\
&&=(x_1,y_1,\ldots,x_{s_k},y_{s_k}).\end{eqnarray*}}
\end{proof}
\begin{theorem}\label{infimum}
Let \sm{\gamma_{i}=(nT_{j'}+j^{''})\alpha_{ki}2\pi\ (1\le i\le s_k)}
and suppose that {\small
\[\begin{array}{r}
C_{jkl2}=C_{jkl2}(X,\sin(\gamma_{1}),\cos(\gamma_{1}),\ldots,
\sin(\gamma_{s_k}),\cos(\gamma_{s_k})).
\end{array}\]} Then {\small \[{\inf\limits_{n\ge1}\{C_{jkl2}(X,\sin(\gamma_{1}),\cos(\gamma_{1}),\ldots,
\sin(\gamma_{s_k}),\cos(\gamma_{s_k}))\}}\]} is equal to
\sm{\min\{C_{jkl2}(X,x_{1},y_{1},\ldots,x_{s_k},y_{s_k})\}} subject
to \sm{\{x_i^2+y_i^2=1,1\le i\le s_k\}.}
\end{theorem}
\begin{proof} According to Lemma \ref{ergodic} and Lemma \ref{infimumL}, it's
sufficient to prove that \sm{T^{s_k}} is the closure of
\sm{\{(e^{\gamma_1\ii},\ldots,e^{\gamma_{s_k}\ii})\}_{n\ge1}.}
Because \sm{\{\alpha_1,\ldots,\alpha_{s_k}\}} are rationally
independent, \sm{\{T_{j'}\alpha_1,\ldots,T_{j'}\alpha_{s_k}\}} are
rationally independent, too. Thus, \sm{T^{s_k}} is the closure of\\
\sm{\{(e^{nT_{j'}\alpha_{k1}2\pi\ii},\ldots,e^{nT_{j'}\alpha_{ks_k}2\pi\ii})\}_{n\ge1}}.
The result of rotating {\small
\[(e^{nT_{j'}\alpha_{k1}2\pi\ii},\ldots,e^{nT_{j'}\alpha_{ks_k}2\pi\ii})\]}
by \sm{(j^{''}\alpha_{k1}2\pi,\ldots,j^{''}\alpha_{ks_k}2\pi)} is
\sm{(e^{\gamma_1\ii},\ldots,e^{\gamma_{s_k}\ii})}. Consequently,
\sm{T^{s_k}} is the closure of
\sm{\{(e^{\gamma_1\ii},\ldots,e^{\gamma_{s_k}\ii})\}_{n\ge1}.}
That completes the proof.
\end{proof}

\subsection{Correctness}

For each \sm{j}, \sm{P_j(X,n)} can be written as {\small
\begin{eqnarray*}
&&D_{j10}(X,n)r_1^n+D_{j11}(X,n)nr_1^n+\ldots+D_{j1d_1}(X,n)n^{d_1}r_1^n+\cdots\\
&&D_{jH0}(X,n)r_{H}^n+D_{jH1}(X,n)nr_{H}^n+\ldots+D_{jHd_{H}}(X,n)n^{d_{H}}
r_{H}^n. \end{eqnarray*}} The \sm{D_{jkl}}'s in the above are real
because \sm{P_{j}(X,n)\in \mathbb{R}} and those \sm{n^lr_k^n}'s are
of different orders. Just like \sm{C_{jkl}}, \sm{D_{jkl}} can be
divided into three parts, {\small
\[D_{jkl}=D_{jkl0}(X)+D_{jkl1}(X,n)+D_{jkl2}(X,n).\]} Because \sm{D_{jkl0}(X)}
contains no \sm{e^{(nT_{j'}+j^{''})\alpha_q2\pi\ii}}'s, the
\sm{e^{(nT_{j'}+j^{''})\alpha_q2\pi\ii}}'s contained in
\sm{D_{jkl1}(X,n)} are periodic and the
\sm{e^{(nT_{j'}+j^{''})\alpha_q2\pi\ii}}'s contained in
\sm{D_{jkl2}(X,n)} are not, \sm{D_{jkl0}, D_{jkl1}} and
\sm{D_{jkl2}} are all real. Since \sm{C_{jkl}(X,n)} results from
\sm{D_{jkl}(X,n)}, we get the following lemma.
\begin{lemma}\label{real0}
For all \sm{n\in\mathbb{N}} and each
\sm{C_{jkl}(X,n)=C_{jkl0}(X)+C_{jkl1}(X)+C_{jkl2}(X,n)},
\sm{C_{jkl0}(X)\in \mathbb{R}}, \sm{C_{jkl1}(X,n)\in\mathbb{R}}
and \sm{C_{jkl2}(X,n)\in\mathbb{R}}.
\end{lemma}

\begin{lemma}\label{real} Consider
\sm{C_{jkl}(X,n)=C_{jkl0}(X)+C_{jkl1}(X)+C_{jkl2}(X,n).}
\begin{enumerate}
\item  \sm{\inf\limits_{n\ge1}\{C_{jkl}(X,n))\}>0}
if and only if {\small
\[\min\{C_{jkl0}(X)+C_{jkl1}(X)+C_{jkl2}(X,y_{11},y_{12},\ldots,y_{s_k1},y_{s_k2})\}>0\]}
subject to \sm{\{y_{t1}^2+y_{t2}^2=1,1\le t\le s_k\}}.
\item If \sm{I=C_{jkl0}(X)+C_{jkl1}(X)+D< 0}, then there is
\sm{c<0} such that \sm{C_{jkl}(X,n)<c} for infinitely many \sm{n}'s,
where {\small
\[D=\min\{C_{jkl2}(X,y_{11},y_{12},\ldots,y_{s_k1},y_{s_k2})\}\]}
subject to \sm{\{y_{t1}^2+y_{t2}^2=1,1\le t\le s_k\}}.
\end{enumerate}
\end{lemma}
\begin{proof}
\begin{enumerate}
\item It has been proved that {\small \[\inf\limits_{n\ge1}
C_{jkl2}(X,n)=\min\{C_{jkl2}(X,y_{11},y_{11},\ldots,y_{s_k1},y_{s_k2})\}\]}
subject to \sm{\{y_{t1}^2+y_{t2}^2=1,1\le t\le s_k\}}. Consequently
{\small
\[\inf\limits_{n\ge1}\{C_{jkl}(X,n))\}=
\min\{C_{jkl0}(X)+C_{jkl1}(X)+C_{jkl2}(X,y_{11},y_{12},\ldots,y_{s_k1},y_{s_k2})\}.\]}
subject to \sm{\{y_{t1}^2+y_{t2}^2=1,1\le t\le s_k\}}.
\item Let \sm{Y=(y_{11},y_{12},\ldots,y_{s_k1},y_{s_k2})} and
\sm{Y'=(y_{11}',y_{12}',\ldots,y_{s_k1}',y_{s_k2}')}. \sm{D} can be
attained because \sm{\{y_{t1}^2+y_{t2}^2=1,1\le t\le s_k\}} is a
bounded closed set and \sm{C_{jkl2}(X,Y)} is a continuous function
of \sm{Y}. Assume that \sm{D=C_{jkl2}(X,Y')}. Since \sm{I< 0} there
exists a neighborhood of \sm{Y'}, say \sm{U}, such that
{\small\[\forall Y\in U\Rightarrow C_{jkl0}(X)+C_{jkl1}(X)+
C_{jkl2}(X,Y)<I/2.\]} Let \sm{c=I/2} and
\sm{\gamma_i=(nT_{j'}+j^{''})\alpha_{ki}2\pi}. Because of the
density of {\small
\[\{(e^{\gamma_1\ii},\ldots,e^{\gamma_{s_k}\ii})\}_{n\ge 1},\]} there
are infinitely many \sm{n}'s such that \sm{(\cos\gamma_1,
\sin\gamma_1,\ldots,\cos\gamma_{s_k},\sin\gamma_{s_k})} lies in
\sm{U}. Thus there are infinitely many \sm{n}'s such that
\sm{C_{jkl}(X,n)<c}.
\end{enumerate}
\end{proof}

If the main algorithm, {\tt Termination}, finds one solution,
\sm{X_0}, the leading coefficient of \sm{G_j(X_0)}, say
\sm{C_{jkl}(X_0,n)}, satisfies \sm{C_{jkl}(X,n)\succ 0}. According
to the definition of \sm{\succ} there exist \sm{c_j>0\
(j=1,\ldots,L)} such that \sm{ C_{jkl}(x,n)>c_j} for all \sm{n}.
Thus \Pone\ is nonterminating. This means that if the algorithm
outputs ``nonterminating", then \Pone\ is nonterminating indeed.

On the other hand, if the main algorithm outputs ``terminating",
then for any \sm{\{C_{jk_jl_j}(X,n),j=1,\ldots,L\}} there is a
subset $V\subseteq \{1,\ldots,L\}$ such that {\small
\[\bigwedge_{j\in V}C_{jk_jl_j}(X,n)\succ 0\]} is not satisfiable
subject to {\small
\[\begin{array}{l} \bigwedge_{j\notin V}C_{jk_jl_j}\succ 0 \wedge
\bigwedge_{(k>k_j) \vee (k=k_j\wedge l>l_j)} C_{jkl}(X,n)\succeq 0
\end{array}\] }

\oomit{{\small
\[\begin{array}{l} \bigwedge_{j\notin V}C_{jk_jl_j}\succ 0 \wedge
\bigwedge_{(k>k_j) \vee (k=k_j\wedge l>l_j)} C_{jkl}(X,n)\succeq 0
\end{array}\] }
is satisfiable but \sm{\bigwedge_{j\in V}C_{jk_jl_j}(X,n)\succ 0}
is not satisfiable.}
According to {\tt the assumption for the main algorithm}, we get
that with the above constraints {\small
\[\forall j\in V,\ \inf\limits_{n\ge 1} {C_{jk_jl_j}(X,n)}\le0.\]} Thus by
Lemma \ref{real}, \sm{\forall j\in V, C_{jk_jl_j}(X,n)} is
identically zero or there are infinitely many \sm{n}'s and some
\sm{c<0} such that \sm{C_{jk_jl_j}(X,n)<c}. 
That means \Pone\ is terminating. Therefore, we get the following
theorem.
\begin{theorem} Under the assumption of the main algorithm,
{\tt Termination} returns ``terminating" if and only if \Pone\ is
terminating.
\end{theorem}
\section{Conjecture}

In this section, we shall discuss the general case of \Pone\ wherein
our assumption for the main algorithm may not hold.

\oomit{Without loss of generality, let's assume that the assignment
matrix \sm{A=Diag(A_1,A_2)} and the linear assignments are
\sm{[X,Y]^T:=Diag(A_1,A_2)[X,Y]^T}, where \sm{A_1} and \sm{A_2} are
two square matrices, the modules of all the eigenvalues of \sm{A_1}
are 1 and the modules of all the eigenvalues of \sm{A_2} are less
than 1. Suppose that the loop guards have the following form
\sm{P_{i1}(X)+P_{i2}(Y)>0,1\le i\le m}. If we add \sm{q(X)^2>0} into
the loop guards, where \sm{q(X)\in \mathbb{Q}[X]}, after
substituting \sm{A_1^nX} for \sm{X} in \sm{q(X)^2}, denote the
resulted \sm{q(X)^2} as \sm{q(X,n)^2} and \sm{P_{i1}(X)} as
\sm{P_{i1}(X,n)}. Let \sm{\diamond\in \{\succ,\succeq\}}.
\sm{S:=\{X:P_{i1}(X)\diamond 0,1\le i\le m\}} is a semi-algebraic
set. If our assumption does not hold and the resulted loop is
nonterminating, there is some \sm{S} such that \sm{\exists X\in S}
and \sm{ \{n:q(X,n)=0\}} contains finitely many elements. However,
\sm{S} may contain only one element, so it is necessary that we can
solve the problem that for any algebraic numbers \sm{X_0}, whether
\sm{\{n:q(X_0,n)\}} contains finitely many elements. Since \sm{X_0}
are algebraic numbers, \sm{q(X_0,n)} can be considered as the
resulted \sm{p(x_{11},x_{12},\ldots,x_{m1},x_{m2})\in\mathbb{A}[X]}
after substituting
\sm{(\sin(n\alpha_12\pi),\cos(n\alpha_12\pi),\ldots,\sin(n\alpha_m2\pi),\cos(n\alpha_m2\pi))}
for \sm{(x_{11},x_{12},\ldots,x_{m1},x_{m2})} in
\sm{p(x_{11},x_{12},\ldots,x_{m1},x_{m2})}. }

Suppose \sm{p(X)=p(x_{11},x_{12},\ldots,x_{m1},x_{m2})\in \QQ[X]},
and one of the loop conditions is \sm{p(X)^2>0}. From the discussion
in Section 3, we know that if we substitute \sm{A^{n+m}X} for \sm{X}
in the conditions, there must be a polynomial \sm{q} such that the
condition becomes \sm{q(X,\sin(n\alpha_12\pi),
\cos(n\alpha_12\pi),\ldots,\sin(n\alpha_m2\pi),\cos(n\alpha_m2\pi))^2>0}.
Because \sm{p} is arbitrary, \sm{q} can be arbitrary. Further,
\sm{\alpha_1,...,\alpha_{m}} can be made rationally independent
because \sm{A} can be arbitrary. It's not hard to see that we can
construct a program \sm{Q} such that it is terminating if and only
if {\small
\[\mathbb{S}_{q,\alpha}\triangleq\{n:q(\sin(n\alpha_12\pi),
\cos(n\alpha_12\pi),\ldots,\sin(n\alpha_m2\pi),\cos(n\alpha_m2\pi))=0\}\]}
contains infinitely many elements.

For any \sm{p(X)\in \mathbb{Q}[X]} the decision problem ``whether
\sm{\{X:p(X)=0\}\bigcap \mathbb{Z}^{2m}=\emptyset}" is undecidable.
\sm{\mathbb{Z}^{2m}} is a ``regular" set while {\small
\[\mathcal{E}=\{(\sin(n\alpha_12\pi),\cos(n\alpha_12\pi),\ldots,\sin(n\alpha_m2\pi),\cos(n\alpha_m2\pi))\}_{n\ge
1}\]} is a chaotic set when \sm{\{\alpha_1,\ldots,\alpha_m\}} are
rationally independent according to the ergodic theory. Intuitively,
deciding whether {\small
\sm{\mathbb{S}_{p,\alpha}=\{X:p(X)=0\}\bigcap\mathcal{E}=\emptyset}}
is more difficult than deciding whether {\small $\{X:p(X)=0\}\bigcap
\mathbb{Z}^{2m}=\emptyset$}. So, we intuitively guess the decision
problem ``whether \sm{\mathbb{S}_{p,\alpha}} is empty" is
undecidable. Following the same idea, we guess the decision problem
``whether \sm{\mathbb{S}_{p,\alpha}} contains infinitely many
elements" is much more difficult and thus undecidable. Thus, we make
the following conjecture:

{\tt Conjecture.} The decision problem ``whether the loop \Pone\ is
terminating over \sm{\mathbb{R}}" is undecidable.

\section{Conclusion}
In this paper we have proved that termination of \Pone\ over
\sm{\mathbb{Z}} is undecidable. Then we give a relatively complete
algorithm, with an assumption, to determine whether \Pone\ is
terminating over \sm{\mathbb{R}}. If the assumption holds, \Pone\ is
terminating iff our algorithm outputs ``terminating". If the
assumption does not hold, \Pone\ is nonterminating if the algorithm
outputs ``nonterminating". We demonstrate the main steps of our
algorithm by an example. Finally we show how hard it is to determine
the termination of \Pone\ by reducing its termination to the problem
of `` whether \sm{\mathbb{S}_{p,\alpha}} has infinite many
elements". We conjecture the latter problem is undecidable. Thus, if
our conjecture holds, the termination of \Pone\ over \sm{\mathbb{R}}
is undecidable.

\section*{Acknowledgement}
The authors would like to thank Prof. Lu Yang for proposing the
problem to us and encouraging us to work on it. We thank Prof.
Michel Waldschmidt for his introduction to Baker's work. We are also
grateful to Prof. Bob Caviness, Prof. Hoon Hong and Prof. Daniel
Richardson for their insightful suggestions. Finally we would like
to thank Prof. Chaochen Zhou, Prof. Naijun Zhan and all the other
members in our group for the discussion at our weekly seminar.

\bibliographystyle{plain}

\begin{thebibliography}{99}
\bibitem{Baker} A. Baker: Linear Forms in the Logarithms of
Algebraic Numbers I, II, III, IV. {\it Mathematika} 13: 204--216,
1966; ibid., 14: 102--107, 220--228, 1967; ibid., 15: 204--216,
1968.

\bibitem{Blon} V.D. Blondel, O. Bournez, P. Koiran, C.H.
Papadimitriou and J.N. Tsitsiklis: Deciding stability and mortality
of piecewise affine dynamical systems. {\it Theoretical Computer
Science}, 255(1--2): 687--696, 2001.

\bibitem{BMS} A.R. Bradley, Z. Manna and H.B. Sipma:
Termination of Polynomial Programs. In {\it Proc. Verification,
Model-Checking, and Abstract-Interpretation (VMCAI)} January 2005,
{\it LNCS 3385}, pp.113--129, 2005.

\bibitem{Mark06}
M. Braverman: Termination of Integer Linear Programs. {\it CAV}
2006, {\it LNCS 4114}, pp.372--385, 2006.

\bibitem{Chen} Y. Chen, B. Xia, L. Yang, N. Zhan and C. Zhou:
Discovering Non-linear ranking functions by Solving Semi-algebraic
Systems. {\it LNCS 4711}, pp.34--49, 2007.

\bibitem{co75}
  G. E. Collins: Quantifier elimination for real closed fields
  by cylindrical algebraic decomposition. In: \emph{Automata Theory and
  Formal Languages} (Brakhage, H., ed.), LNCS {\bf 33}, 134--165.
  Springer, Berlin Heidelberg, 1975.

\bibitem{CH} G. E. Collins and H. Hong: Partial cylindrical algebraic decomposition for quantifier
elimination, {\it Journal of Symbolic Computation}\/, {\bf 12}:
299--328, 1991.

\bibitem{cs01}M. col¡äon and H.B. Sipma: Synthesis of linear ranking functions. In
{\it TACAS¡¯01}, {\it LNCS 2031}, pp.67--81, 2001.

\bibitem{DGG00} D. Dams, R. Gerth, and O. Grumberg: A heuristic for the automatic
generation of ranking functions. In {\it Workshop on Advances in
Verification (WAVe¡¯00)}, pp.1--8, 2000.

\bibitem{Dol} A. Dolzman and T. Sturm: REDLOG: Computer algebra meets computer
logic. {\it ACM SIGSAM Bulletin}\/, {\bf 31}(2): 2--9, 1997.


\bibitem{Mane} R. Mane: {\it Ergodic Theory and Differentiable
Dynamics}. Springer-Verlag, NewYork, 1987.

\bibitem{PR} A. Podelski and A. Rybalchenko: A complete
method for the synthesis of linear ranking functions. In {\it
VMCAI}, {\it LNCS 2937}, pp.465--486, 2004.

\bibitem{Stan97} R. Stanley:
{\it Enumerative Combinatorics}, vol. 1. Cambridge University Press,
1997.

\bibitem{Str}A.W. Strzebonski: Computing in the Field
of Complex Algebraic Numbers. {\it J. Symbolic Computation} 24:
647--656, 1997.

\bibitem{ta51} A. Tarski: \emph{A Decision Method
for Elementary Algebra and Geometry} (2nd edn.). University of
California Press, Berkeley, 1951.

\bibitem{Tiw04} A. Tiwari: Termination of Linear Programs. In
R. Alur and D.A. Peled(Eds.): {\it CAV} 2004, {\it LNCS 3114},
pp.70--82, 2004.

\bibitem{wal}M. Waldschimidt: {\it Diophantine Approximation on Linear Algebraic
Groups}. Springer-Verlag, Berlin, 2000.

\bibitem{xia07}
B. Xia: DISCOVERER: A tool for solving semi-algebraic systems, {\em
Software Demo at ISSAC 2007}, Waterloo, July 30, 2007. Also: {\em
ACM SIGSAM Bulletin}, \textbf{41(3)},102--103, 2007.

\end{thebibliography}

\end{document}